# Towards precise resting-state fMRI biomarkers in psychiatry: synthesizing developments in transdiagnostic research, dimensional models of psychopathology, and normative neurodevelopment

Linden Parkes[1], Theodore D. Satterthwaite[2,3,4], and Danielle S. Bassett[1,2,5,6,7,8]

[1]Department of Bioengineering, School of Engineering & Applied Science, University of Pennsylvania, Philadelphia, PA, 19104 USA.
[2]Department of Psychiatry, Perelman School of Medicine, University of Pennsylvania, Philadelphia, PA 19104, USA.
[3]Lifespan Brain Institute, University of Pennsylvania & Children's Hospital of Philadelphia, Philadelphia, USA
[4] Center for Biomedical Image Computing and Analytics, Perelman School of Medicine, University of Pennsylvania, Philadelphia, PA 19104 USA.
[5]Department of Neurology, Perelman School of Medicine, Philadelphia, PA 19104 USA.
[6]Department of Electrical & Systems Engineering, School of Engineering & Applied Science, University of Pennsylvania, Philadelphia, PA, 19104 USA.
[7]Department of Physics & Astronomy, College of Arts & Sciences, University of Pennsylvania, Philadelphia, PA, 19104 USA.
[8]Santa Fe Institute, Santa Fe, NM 87501 USA

Corresponding author: Danielle S. Bassett, dsb@seas.upenn.edu, Suite 240 Skirkanich Hall, 210 Sth 33rd St, Philadelphia, PA 19104-6321, USA

**Abstract**

Searching for biomarkers has been a chief pursuit of the field of psychiatry. Toward this end, studies have catalogued candidate resting-state biomarkers in nearly all forms of mental disorder. However, it is becoming increasingly clear that these biomarkers lack specificity, limiting their capacity to yield clinical impact. We discuss three avenues of research that are overcoming this limitation: (i) the adoption of transdiagnostic research designs, which involve studying and explicitly comparing multiple disorders from distinct diagnostic axes of psychiatry; (ii) dimensional models of psychopathology that map the full spectrum of symptomatology and that cut across traditional disorder boundaries; and (iii) modeling individuals' unique functional connectomes throughout development. We provide a framework for tying these subfields together that draws on tools from machine learning and network science.

**Introduction**

Precision medicine refers to the idea that diagnosis and treatment strategies for disease processes are optimized when an individual's unique biology is taken into account [1,2]. Critical to this paradigm is the *biomarker*, which is broadly defined as any biological signature that provides an objective indication of an individual's disease status and that ideally predicts clinical outcomes [3]. In recent decades, detecting biomarkers of psychiatric disorders has become a central goal for neuroimaging research as a means to drive psychiatry towards precision medicine [1]. To this end, a large body of literature has emerged using resting-state functional magnetic resonance imaging (rs-fMRI) to characterize the brain dysfunction accompanying psychiatric conditions [4]. However, a combination of high symptomatic and biological heterogeneity within disorders as well as comorbidity amongst disorders [5,6] has resulted in a lack of disorder specificity in candidate biomarkers. This lack of specificity has rendered the field unable to translate knowledge reliably into clinical practice [7]. Here, we discuss three recent developments in rs-fMRI research that hold promise for improving our capacity to generate biomarkers for mental health. First, we highlight a shift in focus from studying single disorders in isolation to transdiagnostic paradigms that examine multiple disorders in a single study. Second, we discuss insights from studies that take a dimensional, as opposed to categorical, approach to measuring psychopathology. Third, we briefly review how understanding the developing functional connectome is providing new directions into studying brain dysfunction. Finally, we provide a framework for tying these three areas of research together drawing on machine learning models of normative neurodevelopment [8,9] and tools from network science [10,11].

1. **A shift toward transdiagnostic research**

The standard paradigm for biomarker detection in psychiatry is the case-control design, a categorical approach wherein *case* refers to a group of individuals experiencing a given mental disorder, and *control* refers to a group of individuals not experiencing the disorder [12]. Typically, individuals in the *control* group are selected because they are not experiencing any mental disorder (i.e., *healthy control*), hence any observed group differences are attributed to the disease process. Since the inception of the rs-fMRI paradigm [13], case-control rs-fMRI studies have been conducted in almost every major disorder described by psychiatric nosologies (e.g., DSM-5, ICD-11). However, within a single study, the case-control paradigm is typically applied to a single disorder; rarely are multiple disorders studied concurrently. Consequently, recent meta-analytic work has revealed that many of the field's candidate biomarkers lack disorder specificity. Sha *et al*. [14] examined 182 rs-fMRI studies spanning attention deficit hyperactivity disorder (ADHD), autism spectrum disorder, bipolar disorder, depression, obsessive-compulsive disorder (OCD), post-traumatic stress disorder, and schizophrenia (as well as several neurological disorders) and found that dysfunction in the default-mode, sensorimotor, fronto-parietal, and subcortical systems was common to all disorders. This result challenges the idea that large-scale dysconnectivity between these systems represents a disorder-specific biomarker (see Dong *et al*. [15] for a recent example in schizophrenia). Instead, this result demonstrates that in order to attain specificity, the field needs to transition to transdiagnostic designs, wherein multiple disorders are examined concurrently, and differential diagnostic sensitivity can be directly assessed.

One candidate biomarker where progress toward disorder specificity has shown recent promise is dysfunction in the frontostriatal circuits [16]. Frontostriatal circuits are a set of parallel circuits that topographically connect different regions of the frontal cortex to different subregions of the striatum [17,18]. Using a combination of resting-state activity within the striatum, as well as intra- and extra- striatal connectivity, Li *et al.* [16] trained a support vector machine (SVM) to separate schizophrenia patients from healthy controls. Then, the authors took the distance between each individual and the separating SVM hyperplane as an index for frontostriatal abnormality (FSA). Next, they compared FSA scores from individuals with schizophrenia, bipolar disorder, depression, OCD, and ADHD against healthy controls and found that *only* the schizophrenia and bipolar groups showed significant differences in FSA scores, with differences being greater for schizophrenia relative to bipolar. This pattern of results suggests that the author's SVM learned multivariate patterns of frontostriatal dysfunction that *selectively* separated schizophrenia and bipolar from healthy controls. However, such selectivity does not imply that disorder-specific frontostriatal dysfunction is absent in other disorders (e.g., OCD [19,20]). Li *et al*. trained their SVM to separate schizophrenia and healthy controls, and then used that model to generate FSA scores for other disorder groups. Thus, a different SVM could be trained to separate OCD from healthy controls and to test for selective frontostriatal dysfunction in OCD. Further, this OCD-trained SVM could be compared to the schizophrenia-trained SVM to characterize patterns of frontostriatal dysfunction that support differential diagnosis.

2. **Dimensions of psychopathology**

While transdiagnostic case-control designs may assist in identifying which biomarkers are disorder-specific or disorder-general, these study designs remain focused on the group-level signatures revealed by comparing healthy individuals with those that meet criteria for a mental disorder. Consequently, they fail to capture subclinical variation in symptomatology that is important for understanding the link between brain dysfunction and psychiatry [21,22]. In response to this limitation, rs-fMRI studies have begun taking an explicitly dimensional approach, wherein the full range of variation -- normal, subclinical, and abnormal -- in symptomatology and brain function is characterized.

Recent rs-fMRI studies deploying dimensional models have varied in terms of the number and scope of the estimated dimensions of psychopathology. On the one hand, some studies have aimed to quantify dimensions of psychopathology that cut across a broad range of diagnostic entities, with a view to teasing apart which biomarkers are unique to, or common across, most diagnostic axes of psychiatry. These studies have often drawn on the idea of the *p-factor* [23,24], a statistical concept that posits that the endorsement of any psychiatric symptom increases the probability of endorsing any other symptom (so-called *overall psychopathology*). Studies typically estimate the *p-factor* alongside multiple specific dimensions of psychopathology (e.g., depression, psychosis) in a hierarchical fashion, rendering the specific dimensions orthogonal to each other and the *p-factor.* This model structure allows researchers to uncover both disorder-general and disorder-specific biomarkers. On the other hand, some studies have focused on decomposing a single diagnostic axis of psychiatry (e.g., psychosis) into multiple subcomponents (e.g., positive and negative psychosis dimensions), aiming instead to discover multiple distinct biomarkers within select disorders.

At the multi-diagnosis level, recent studies by Elliot *et al*. [25] and Kebets *et al*. [26] found that greater scores on the *p-factor* were associated with hypo-connectivity within sensory and somatomotor systems, as well as with hyper-connectivity between these systems and the executive systems. These results point toward a disorder-general biomarker of dysconnectivity between lower- and higher-order systems in the cortical hierarchy. Notably, convergence between these studies was present despite the use of distinct statistical methods (for a review of the methods see Kaczkurkin *et al*. [27]). Briefly, Elliot *et al.* modeled latent psychopathology dimensions independently of the rs-fMRI data (i.e., a so-called single-view approach), and then related the *p-factor* to functional connectivity. By contrast, Kebets *et al*. deployed a multi-view approach that jointly estimated latent dimensions of psychopathology and functional connectivity in a single model and labeled one of their dimensions as a *p-factor post hoc*. However, another multi-view study by Xia *et al*. [28] found instead that dysconnectivity between the higher-order default mode and executive systems was common across dimensions of mood, psychosis, fear, and externalizing behavior. That Xia *et al*.'s results implicated the higher-order cortical systems while Kebets *et al*. and Elliot *et al*.'s results implicated the lower-order cortical systems may be due to the fact that Xia *et al*. did not find a *p-factor* in their data*.* Instead, Xia *et al.* derived disorder-general biomarkers by examining the spatial correspondence between patterns of dysconnectivity observed for their specific dimensions. This pattern of results suggests that the presence or absence of a *p-factor* impacts the ensuing biomarkers. Indeed, how to best model broad dimensions of psychopathology that span multiple diagnostic categories remains open to debate in the field [29].

At the single-diagnosis level, dimensional models have been used to uncover differences in functional connectivity across subdimensions of specific forms of psychopathology. Sabaroedin *et al*. [30] examined the extent to which normal and subclinical variation in negative and positive psychosis-like experiences (PLE) tracked differences in resting-state frontostriatal connectivity in a large population cohort. Convergent with previous literature in schizophrenia, greater positive PLEs were associated with reduced connectivity between the dorsal striatum and prefrontal cortex, demonstrating that frontostriatal dysfunction in schizophrenia may generalize to subclinical levels of psychotic phenomenology. Taking a dimensional and transdiagnostic approach, Parkes *et al*. [19] examined frontostriatal dysfunction in individuals with OCD and individuals with gambling disorder (GD). The authors examined whether dimensional measures of impulsivity and compulsivity, two constructs with joint relevance to OCD and GD, tracked variation in resting-state effective connectivity in frontostriatal circuits. Compared to pairwise case-control and case-case comparisons, Parkes *et al*. found that inter-individual differences in compulsivity better explained frontostriatal dysconnectivity when patients and controls were pooled into a single group. These studies show that splitting specific dimensions of psychopathology into subdimensions reveal addition insights in brain dysfunction and, together with the broad multi-diagnosis studies, highlight the importance of considering dimensional models of psychopathology at multiple scales of specificity.

### 3. Modeling the connectome throughout development

In addition to transdiagnostic and dimensional designs, linking mental disorder to rs-fMRI has also been facilitated by progress in modeling the connectome throughout development. In a connectome *identifiability* [31] study (**Box 1**), Kaufmann *et al*. [32] showed that individuals' whole-brain connectomes became increasingly distinct from one another throughout childhood and adolescence, demonstrating that the whole-brain connectome encodes individualized fingerprints that emerge with age. Critically, individuals with greater scores on the *p-factor* showed reduced connectome distinctiveness compared to healthy individuals, indicating that disruptions to the formation of these fingerprints is relevant to mental health. Subsequent longitudinal work showed that whole-brain connectomes could be used to identify participants across time intervals of several months and up to 2 years [33,34], demonstrating that connectome fingerprints are stable over long periods.

The above studies have also illustrated that different brain systems have a differential impact on identifiability. Horien *et al*. [33] and Miranda-Dominguez *et al*. [34] each found, across multiple datasets, that higher-order executive systems offered higher identifiability than lower-order sensory and motor systems; the latter only outperformed subcortical systems [33]. This differential system effect remained throughout childhood and adolescence, and into adulthood [32,35]. Thus, whereas higher-order systems develop into increasingly individualized functional fingerprints, lower-order systems remain stable and more homogenous across individuals. Critically, Kaufmann *et al*. [32] found that the reduced connectome distinctiveness associated with greater *p-factor* scores was more pronounced in the executive systems, suggesting that disruptions to an individual's higher-order functional fingerprint is indicative of mental disorder.

The above neurodevelopmental work stands in contrast to the studies demonstrating that greater *p-factor* scores were predominantly associated with dysconnectivity centered on the lower-order sensory and somatomotor systems [25,26]. This discrepancy may be explained by differences between edge-level modeling and distance-based modeling of inter-individual differences in the connectome. For example, Kebets *et al*. [26] deployed an edge-level approach testing whether dimensions of psychopathology covaried directly with edge strength and found that the strongest signals originated from lower-order cortical systems. In contrast, identifiability studies effectively score individuals based on the distance between their whole-brain connectomes across scans in an *N*-dimensional space (where *N* denotes the number of edges). Subsequently, covariation with psychopathology indexes inter-individual differences in the extent to which connectomes are individualized, signals for which appear concentrated in higher-order systems [32,33]. This finding suggests that mental disorders affect different parts of the cortical hierarchy in qualitatively distinct ways, with edge-level modeling being sensitive to lower-order systems that are relatively stable and homogenous across people, and the distance-based modeling being sensitive to higher-order systems that are highly and increasingly individualized throughout development.

**Future directions**

The intersection between transdiagnostic research, dimensional models of psychopathology, and connectome development has the potential to uncover novel biomarkers in psychiatry. Synthesizing these approaches will require the joint application of tools from machine learning and network neuroscience [36]. In particular, recent developments in machine learning have provided new tools for modeling, and detecting abnormalities in, neurodevelopment [8]. This approach, known as *normative modeling*, involves using Bayesian regression techniques to model the non-linear age-related changes in neuroimaging data in healthy individuals. Importantly, these techniques provide estimates of percentiles of variation within the healthy cohort [9]. In contrast to case-control designs, which are restricted to estimating brain abnormalities at the group level, these percentiles of variation allow researchers to quantify the extent to which patients deviate from the normative age trajectory on an individual basis. In psychiatry research, normative models have already been used to illustrate the large amount of heterogeneity in brain structure present within disorders such as schizophrenia [37], bipolar [37], ADHD [38], and autism [39,40]. These studies have shown that few patients, if any, have spatially consistent patterns of age-related brain abnormalities, supporting the notion that case-control paradigms provide little utility in biomarker detection.

The extant normative modeling literature has so far focused on indices of brain structure, in part because normative models depend upon a robust underlying age effect in order to yield interpretable abnormalities. Robust regional age effects are trivially obtained using structural indices [41,42] (e.g., gray matter volume), but this is not true of edge-level estimates of resting-state functional connectivity. Given the age effects reported in the identifiability literature [32,43], normative models may be well suited to identifying individual patients with neurodevelopmentally delayed (i.e., reduced) connectome individualization (**Figure 1**). Using pairs of rs-fMRI (or task-based fMRI) scans from longitudinal developmental cohorts (**Figure 1A**), normative models of connectome individualization can be estimated in typically developing individuals (**Figure 1B**, ●). Next, deviations can be estimated in a separate cohort (**Figure 1B**, ●) and correlated with dimensional psychopathology models. Such models may assist in identifying which axes of psychiatry are associated with the abnormal formation of individual connectomes, which may be underpinned by disorganization in higher-order brain systems.

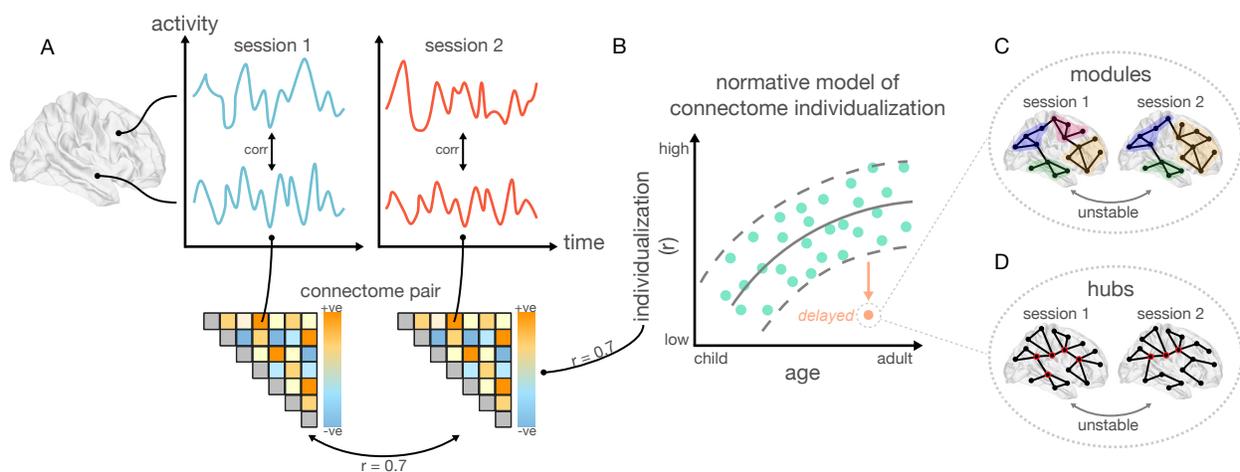

**Figure 1. Abnormally delayed connectome individualization may be characterized using a combination of normative modeling and tools from network science. A**, For each individual, pairs of whole-brain resting-state connectomes may be obtained from different scan sessions (〰️, 〰️). Correlations between connectome-pairs may be used to index individualization (e.g., *r* = 0.7). **B**, Normative models of age-related increases in individualization (●) can then be specified, and the individual's deviating connectomes (●) can be further characterized using tools from network science (**C, D**).

The issue of biological heterogeneity in psychiatry reveals a clear 'many to one' problem, where multiple pathophysiological pathways converge on similar forms of psychopathology. Thus, normative models of connectome individualization may give rise to individuals with quantitatively similar deviations in individualization, and levels of psychopathology, that correspond to qualitatively distinct underlying disruptions in their connectome architecture. We anticipate that tools from network science will be critical to uncovering these unique disruptions [44,45]. For example, brain regions in the connectome are organized into functional modules [46] -- characterized by denser connectivity within modules than between modules -- and abnormal individualization may be reflected in the instability of an individual's modular structure across scans (**Figure 1C**). Similarly, hub nodes [47,48] -- brain regions whose densely connected nature are thought to support the integration of information across functional modules -- may be less stable across an individual's scans (**Figure 1D**). These examples are not exhaustive and we note that many other avenues may also be of interest, including developments in the areas of time-varying functional connectivity [49,50] and individual-specific parcellations [51–54]. Critically, the nature of an individual's connectome instability, as indexed by network science, need not be the same across individuals with similarly high scores on dimensions of psychopathology, accommodating the notion that unique patterns of pathophysiology give rise to similar symptom profiles.

**Conclusions**

While initially fruitful, single disorder case-control studies in psychiatry are unlikely to continue to yield advances in biomarker research in the field of psychiatry. Here, we briefly discussed recent rs-fMRI work in transdiagnostic psychiatry, dimensional psychopathology, and neurodevelopment that we believe are driving the field towards more precise biomarkers in mental health. While some differences in methodologies remain to be thoroughly examined, results suggest that edge-level and distance-based models (i.e., identifiability) yield complementary information relevant to mental health. We suggest that combining normative models of connectome individualization with transdiagnostic dimensional models of psychopathology and tools from network science hold great promise for understanding the link between brain dysfunction and mental disorder.

**Box 1 Individuals have unique connectome fingerprints**

Studying individual differences in resting-state connectivity has been aided by a shift away from the mass-univariate study of edge-level functional connectivity towards studying the whole-brain connectome as a single multivariate object. In this context, the brain is represented as a symmetric *N*×*N* adjacency matrix, **A**, where *N* represents a number of discrete brain regions referred to as nodes. Within this matrix, elements $A_{ij}$ take on a weighted value corresponding to the functional connectivity between nodes *i* and *j*. This weight can be estimated by a simple Pearson correlation coefficient between rs-fMRI timeseries, or by other more sophisticated measures [55]. One promising approach for analyzing multivariate inter-individual differences in the connectome is known as *identifiability*. Popularized by Finn *et al*. [31], identifiability involves correlating connectivity estimates within pairs of whole-brain connectomes. The simplest setup involves generating at least two whole-brain connectomes for each individual in a sample (e.g., across multiple scans/sessions) and then calculating the correlation between all possible connectome pairs in a sample. Multiple studies have found that correlation coefficients are higher for connectome pairs taken from the same individual than for connectome pairs taken across different individuals, allowing for the identification of individuals over repeated scans with near perfect accuracy [31,56,57]. This phenomenon illustrates that the multivariate pattern of whole brain functional connectivity is highly unique to each individual.

**Citation diversity statement**
Recent work in neuroscience and other fields has identified a bias in citation practices such that papers from women and other minorities are under-cited relative to the number of such papers in the field [58–61]. Here we sought to proactively consider choosing references that reflect the diversity of the field in thought, form of contribution, gender, and other factors. We used automatic classification of gender based on the first names of the first and last authors [58,62], with possible combinations including male/male, male/female, female/male, and female/female. Excluding self-citations to the first and last authors of our current paper, the references contain 58.5% male/male, 12.2% male/female, 19.5% female/male, 4.9% female/female, and 4.9% unknown categorization. We look forward to future work that could help us to better understand how to support equitable practices in science.

**Acknowledgements**

LP, TDS, and DSB primarily acknowledge financial support from the National Institute of Mental Health through R01 award MH113550 to TDS and DSB. Additional support was provided by the National Institutes of Mental Health through awards 1R01MH119219 (PI: Gur) and 1R01MH120482-01 (PI: TDS), and the Army Research Laboratory through CPNF Subcontract PO #182200 (PI: DSB). The authors acknowledge Dr. Eli J. Cornblath, Dr. Arun Mahadevan, and Mr. Dale Zhou for valuable discussions during the writing of the manuscript.

**Highlighted References:**

(**) Sha, Z. *et al.* Meta-Connectomic Analysis Reveals Commonly Disrupted Functional Architectures in Network Modules and Connectors across Brain Disorders. *Cerebral Cortex* **28**, 4179–4194 (2018).

- This study performed a comprehensive meta-analysis of rs-fMRI connectome data and showed that patterns of dysfunction were common across multiple psychiatric populations.

(**) Li, A. *et al.* A neuroimaging biomarker for striatal dysfunction in schizophrenia. *Nat Med* **26**, 558–565 (2020).

- This study trained a support vector machine to separate patients with schizophrenia from healthy controls and showed that their model did not generalize to other psychiatric groups, demonstrating disorder-specificity.

(**) Kebets, V. *et al.* Somatosensory-Motor Dysconnectivity Spans Multiple Transdiagnostic Dimensions of Psychopathology. *Biological Psychiatry* **86**, 779–791 (2019).

- This study used partial least squares to demonstrate that dysconnectivity within lower-order cortical systems and between lower- and higher-order systems may be a common biomarker across many mental disorders.

(**) Horien, C., Shen, X., Scheinost, D. & Constable, R. T. The individual functional connectome is unique and stable over months to years. *NeuroImage* **189**, 676–687 (2019).

- This study demonstrated that individual's unique functional fingerprints were stable over months to years, and that higher-order systems drove this uniqueness to a greater extent than lower-order systems.

(*) Elliott, M. L., Romer, A., Knodt, A. R. & Hariri, A. R. A Connectome-wide Functional Signature of Transdiagnostic Risk for Mental Illness. *Biological Psychiatry* **84**, 452–459 (2018).

- This study combined a dimensional model of psychopathology with connectome-wide association analysis to demonstrate that dysconnectivity

between the visual cortex and higher-order executive systems was associated with greater scores on the *p-factor.*

(*) Kaufmann, T. *et al.* Delayed stabilization and individualization in connectome development are related to psychiatric disorders. *Nature Neuroscience* **20**, 513–515 (2017).

- This study showed that abnormal patterns of connectome distinctiveness throughout development was common to multiple mental disorders, and that this effect was concentrated in higher-order systems.

(*) Marquand, A. F. *et al.* Conceptualizing mental disorders as deviations from normative functioning. *Mol Psychiatry* (2019).

- This article discusses the concept of identifying abnormal brain signatures on an individual rather than a group basis. It provides an overview of the normative modeling methodology and how this approach will help to disentangle biological heterogeneity in psychiatry.